\documentclass[twocolumn,showpacs,preprintnumbers,pra,aps,amssymb,amsfonts,floatfix]{revtex4}
\usepackage{times,graphicx}
\newcommand{\trace}{\mathop{\rm Tr}\nolimits}

\newcommand{\bra}[1]{\langle#1|}
\newcommand{\ket}[1]{|#1\rangle}

\newcommand{\qed}{\hfill$\square$\par\vskip24pt}

\newcommand{\cP}{{\cal P}}

\newcommand{\identity}{{\text{\rm\openone}}} 
\newcommand{\id}{{\text{\rm\openone}}} 

\newcommand{\be}{\begin{equation}}
\newcommand{\ee}{\end{equation}}
\newcommand{\bea}{\begin{eqnarray}}
\newcommand{\eea}{\end{eqnarray}}
\newcommand{\beas}{\begin{eqnarray*}}
\newcommand{\eeas}{\end{eqnarray*}}

\newtheorem{theorem}{Theorem}

\newcount\minute
\newcount\hour
\def\currenttime{%
    \minute\time
    \hour\minute
    \divide\hour60
    \the\hour:\multiply\hour60\advance\minute-\hour\the\minute}
\begin{document}
\title{Entanglement on mixed stabiliser states: Normal Forms
and Reduction Procedures}
\author{Koenraad M.R. Audenaert and Martin B. Plenio}
\affiliation{Blackett Laboratory, Imperial College London, Prince
Consort Road, London SW7 2BW, United Kingdom}
\affiliation{Institute for Mathematical Sciences, Imperial College
London, Exhibition Road, London SW7 2BW, United Kingdom}
\date{\today, \currenttime}
\begin{abstract}
The stabiliser formalism allows the efficient description of a
sizeable class of pure as well as mixed quantum states of
$n$-qubit systems. That same formalism has important applications
in the field of quantum error correcting codes, where mixed stabiliser states
correspond to projectors on subspaces associated with stabiliser codes.

In this paper, we derive efficient reduction
procedures to obtain various useful normal forms for stabiliser
states. We explicitly prove that these procedures will always
converge to the correct result and that these procedures are
efficient in that they only require a polynomial number of
operations on the generators of the stabilisers.

On one hand, we obtain two single-party normal forms. The first,
the row-reduced echelon form, is obtained using only permutations
and multiplications of generators. This form is useful to
calculate partial traces of stabiliser states. The second is the
fully reduced form, where the reduction procedure invokes
single-qubit operations and CNOT operations as well. This normal
form allows for the efficient calculation of the overlap between
two stabiliser states, as well as of the Uhlmann
fidelity between them, and their Bures distance.

On the other hand, we also find a reduction procedure of bipartite
stabiliser states, where the operations involved are restricted to
be local ones. The two-party normal form thus obtained lies bare a
very simple bipartite entanglement structure of stabiliser states.
To wit, we prove that every bipartite mixed stabiliser state is
locally equivalent to a direct product of a number of maximally
entangled states and, potentially, a separable state. As a
consequence, using this normal form we can efficiently calculate
every reasonable bipartite entanglement measure of mixed
stabiliser states.
\end{abstract}
\maketitle
\section{Introduction}
The exploration of the properties of quantum entanglement is one
of the main branches of quantum information theory \cite{Plenio V
98,Eisert P 03,Plenio V 05}. While a reasonably detailed
understanding of two-qubit entanglement has been achieved, the
entanglement properties of higher-dimensional or multi-particle systems
remain largely unexplored, with only
isolated results \cite{Linden PSW 99}. This is largely due to the
complexity involved in these investigations, which in turn
originates from the tensor product structure of the multi-particle
state space. This structure leads to an exponential growth in the
number of parameters that are required for the description of the
state. The same problem arises when one attempts to consider the
time-evolution of a many-body quantum system or, say, of a quantum
computation. Generally, a significant part of the Hilbert space
and consequently an exponential number of
parameters are required to describe the quantum system at all
times.

One way of approaching this situation is to impose constraints on
the set of states and/or the set of operations that one is
interested in without curtailing the variety of possible
qualitative entanglement structures too much. In this context an
interesting class of states that arises is that of stabiliser
states \cite{Gottesman 97,Gottesman 98,Briegel R 00,Nielsen C
00,Hostens DM 04,VandenNest DM 04a} which, via the concept of
graph states \cite{Schlingemann 01,Schlingemann 03,Hein EB
03,VandenNest DM 04f,Guehne THB 04,Dur AB 03,VandenNest DM
04d,VandenNest DM 04c} have some connection with graph theory. The
feature of these states that allows for a more detailed study of
their entanglement properties is the fact that an n-particle
stabiliser state is determined as the joint unique eigenvector with
eigenvalue $+1$ of a set of only $n$ tensor products of Pauli
operators. This results in a very compact description of the
quantum state requiring only of order $O(n^2)$ parameters and
therefore provides hope that a more detailed understanding of
their entanglement structure can be obtained. Despite this
simplification, stabiliser states still exhibit multi-particle
entanglement and permit, for example, the violation of local realism
\cite{Guehne THB 04}.

The stabiliser formalism not only allows for the efficient
description of a certain type of quantum states, but also permits
the efficient simulation of a restricted, but nevertheless
interesting, class of time evolutions, namely those that respect
the tensor product structure of the Pauli operators \cite{Anders B
05,Calsamiglia HDB 05,Aaronson G 04,Schlingemann 02}. Again, these
simulations can be performed in polynomial time in the number of
qubits, in stark contrast to the simulation of a general time
evolution of an $n$-qubit system, which requires an amount of
resources that is exponential in $n$.

The stabiliser formalism uniquely specifies the quantum state of
an $n$-qubit system employing only polynomial resources. This
alone, however, is not sufficient. It is also important to be able
to derive {\em all} relevant physical quantities, especially those
relating to entanglement, directly from the stabiliser formalism.
Indeed, having first to deduce the state explicitly and then
computing the property from the state would generally involve an
undesirable exponential overhead in resources. While one can
expect a direct approach to be possible in principle, it is
evident that detailed and explicit presentations of algorithms to
achieve these tasks in a systematic way and whose convergence is
proven are of significant interest. In the context of entanglement
properties some effort has recently been expended in this
direction in \cite{Fattal CYBC 04}, where, employing sophisticated
tools from group theory, the existence of a useful entanglement
measure for multi-particle stabiliser states was demonstrated.

The present work progresses further in a similar direction.
Employing elementary tools we present a number of normal forms for
pure {\em and} mixed stabiliser states, together with explicit and
detailed descriptions of algorithms (including proofs of
convergence) that allow the reduction to these normal forms. In
turn, these normal forms then permit us to compute {\em any}
entanglement measure, overlaps between stabiliser states and
various other quantities. Detailed descriptions of the algorithms
are provided that should make it straightforward to implement
these algorithms in any programming environment and we are able to
provide a ($\beta$-tested) suite of MatLab programs on request.

This suite can then form the basis for more detailed studies and
further applications of the stabiliser formalism to a whole range
of physical questions (see also \cite{Calsamiglia HDB 05}). This
will be reported on in a forthcoming publication.

The stabiliser formalism also plays a central role in the field of
quantum error correcting codes. Mixed stabiliser states (defined in Section II)
are in one-to-one correspondence with projectors on subspaces associated with
stabiliser codes \cite{Gottesman 97}. Although our normal forms and reduction procedures
have been designed with applications to entanglement theory in mind, they might
have a bearing on stabiliser codes as well.

The present paper is organized as follows: In Section II the basic
notations and conventions we use are introduced while Section III
describes the elementary operations that will form the basis of
all reduction procedures.

The single-party normal forms are the topic of Sections IV, V and
VI. Section IV deals with the so-called row-reduced echelon form
(algorithm RREF), the reduction to which is based on row
operations only. It allows to check independence of any (putative)
set of generators and to calculate partial traces (algorithm
PTRACE). In Section V we describe the full reduction procedure
(algorithm CNF1) to single-party normal form, using row and column
(qubit) operations. In Section VI we present an algorithm
(algorithm OVERLAP) that is based on the full reduction and allows
to calculate overlaps between stabiliser states, Uhlmann fidelity, and Bures distance.

In Section VII we turn to the bipartite case, where we prove that
the bipartite entanglement structure of stabiliser states is
remarkably simple. To wit, we show that mixed bipartite stabiliser
states are locally equivalent to a tensor product of a certain
number of maximally entangled 2-qubit states and, potentially, a
fully sparable mixed state. We present an algorithm (algorithm
CNFP) to obtain the number of these maximally entangled pairs,
allowing for the calculation of any reasonable bipartite
entanglement measure.

We conclude the description of our findings in Section VIII.
\section{Notations and Conventions}
A {\em stabiliser operator} on $N$ qubits is a tensor product of
operators taken from the set of Pauli operators
\be
X:=\left(\begin{array}{cc}0&1\\1&0\end{array}\right),\,\,
Y:=\left(\begin{array}{cc}0&-i\\i&0\end{array}\right),\,\,
Z:=\left(\begin{array}{cc}1&0\\0&-1\end{array}\right),
\ee
and the identity $\id$.
An example for $N=3$ would be the operator
$g=X\otimes\id\otimes Z$. A set $G=\{g_1,\ldots,g_K\}$ of $K$
mutually commuting stabiliser operators that are independent,
i.e.\ $\prod_{i=1}^{K} g_i^{s_i}=\id$ exactly if all $s_i$ are
even, are called the {\em generator set} for the {\em stabiliser
group} $S$. This stabiliser group $S$ then consists of
all distinct products of operators from the generator set.

For $K=N$ a generator set $G$ uniquely determines a single state
$|\psi\rangle$ that satisfies $g_k|\psi\rangle=|\psi\rangle$ for
all $k=1,\ldots,N$. Any state for which such a generator set
exists is called {\em stabiliser state}. Such a state has
trivially the property that
$g_k|\psi\rangle\langle\psi|=|\psi\rangle\langle\psi|$ for all $k$
so that
\begin{equation}
    |\psi\rangle\langle\psi| =
    \frac{1}{2^N} \sum_{g\in S} g. \label{eq:pureorig}
\end{equation}
This formula depends on the complete set of stabilisers, and is, therefore, not very practical.
The following formula expresses the state in terms of a generator set \cite{Nielsen C 00}
\begin{equation}
    |\psi\rangle\langle\psi| = \prod_{k=1}^N \frac{\id+g_k}{2}. \label{eq:projpure}
\end{equation}
The procedure presented in Section \ref{sec:RREF} yields as a side
result an elementary proof of this statement.

Considering two parties $A$ and $B$,
the reduced density matrix of the stabiliser state can be computed as
\begin{equation}\label{eq:mixedorig}
    \rho_A = \trace_B\rho = \frac{1}{2^N} \sum_{g\in S} \trace_B g.
\end{equation}
This obviously means that only operators $g$ contribute that have
identity operators acting on all qubits belonging to B. Needless
to say, computing $\rho_A$ from $\rho$ directly is hopelessly
inefficient; as there are $2^N$ different $g$, this task requires
an exponential number of operations in general.

It turns out, however, that there is a class of mixed states that
can also be characterised employing stabilisers. We will call
these \textit{mixed stabiliser states}, and they contain the pure
stabiliser states as a subset. The important feature of this class
is that the reduced density matrix of a mixed stabiliser state is
again a mixed stabiliser state. Furthermore, as we will show below
in Section \ref{sec:ptrace}, the stabiliser group of a reduction
can be efficiently calculated directly from the original
stabiliser group, without calculating the state and its reduction
explicitly.

To characterise mixed stabiliser states, one simply considers sets $G$
that are linearly dependent. As a consequence, by multiplying
stabilisers, one can achieve that some of them become identical to
$\id$ and only $K$ linearly independent ones remain. Then
the common eigenspace of these $K$ operators will have a dimension
larger than $1$. As in eq.~(\ref{eq:projpure}) one immediately
deduces that the density operator is again just the projector onto
this eigenspace, rescaled to trace 1:
\bea
    \cP &=& \prod_{k=1}^K \frac{\id+g_k}{2} \nonumber \\
    \rho &=& \frac{1}{2^{N-K}}\cP.
\label{eq:projmixed}
\eea
Given that $\cP$ is a projector onto a subspace of
dimension $2^{N-K}$, the entropy of $\rho$ is simply $N-K$.
In analogy with matrix analysis terminology, we will call $K$ the \textit{rank} of the stabiliser group.
Stabiliser groups with $K=N$ will be called \textit{full-rank},
and stabiliser groups with $K<N$ \textit{rank-deficient}.

\medskip
\textbf{Note:} In case of a rank-deficient stabiliser group one
has to distinguish between \textit{stabiliser} states and
\textit{stabilised} states. The stabiliser state is the one given
by (\ref{eq:projmixed}), and the stabiliser formalism allows to
study its properties in an efficient way. On the other hand, there
are many states that are stabilised by that same stabiliser group,
but in general they are not stabiliser states. Indeed, most of
these stabilised states cannot be described as ``the unique state
stabilised by a full-rank stabiliser group'', and hence, the
stabiliser formalism cannot be used to study their properties via
that group. For example, any state is stabilised by the
(singleton) stabiliser group $\{\id\}$, but only the maximally
mixed state $\id/2$ is the stabiliser state corresponding to that
group.

\medskip
For the following the aim will be to derive basic entanglement
properties such as the entropy of entanglement or the logarithmic
negativity for stabiliser states, pure or mixed, directly from
their generating set. To this end it will be useful to find a
normal form for the generator set that reveals the relevant
entanglement structure.


We now introduce the concept of \textit{stabiliser array}, which is quite simply
a rectangular array of $K$ rows and $N$ columns, where the elements are Pauli matrices or the Identity
matrix.
Specifically, the element in the $k$-th row and $n$-th column of the stabiliser array
corresponding to a generator set $G=\{g_1,\ldots,g_K\}$ on $N$ qubits
is the $n$-th tensor factor (corresponding to qubit $n$)
of the $k$-th generator $g_k$.
For some applications it will also be necessary to deal with the generator phase factors.
While in general these phase factors can assume the values $\pm1$ and $\pm i$,
for the purpose of describing stabiliser states only the values $\pm1$ make sense (since states are Hermitian).
We will store these phase factors in a $K$-dimensional vector $s$, where $s_k$ is the phase factor of generator $g_k$.

The purpose of the various normal forms that will be presented in this paper are
to structure the set of stabiliser states into certain equivalence classes.
They are similar in spirit to the normal forms that have been devised for matrices.
For example, the row-reduced echelon form, which exhibits the rank of a matrix, has a counterpart for stabiliser arrays.
Despite this similarity, the normal forms presented here are of an entirely different nature.
The rank of a stabiliser array, which we will introduce in Section \ref{sec:RREF}, is akin to the rank
of a matrix in that it equals the number of independent generators in a generator set,
but there the similarity stops. In linear algebra one defines both row and column rank of a matrix
and one proves that these two ranks are actually equal. For stabiliser arrays, one cannot even give a meaningful
definition of column rank.

These differences ultimately boil down to the fact that a stabiliser array is not really a matrix.
The two foremost reasons are that its elements are not numbers but elements of the Pauli group, and
second, that matrices represent linear operations in linear spaces, while
stabiliser arrays represent sets (namely, sets of generators).
As a consequence, while operations like matrix transpose, matrix multiplication, addition, and inverse
make perfect sense for matrices, they are utterly meaningless for stabiliser arrays.
The allowed operations on stabiliser arrays are thus much more
restricted than in the matrix case. For example, the only row operations that make sense for stabiliser arrays
are row interchange and elementwise row multiplication (which is based on the Pauli group multiplication law).
This will be discussed in more detail in the following Section.
This explains the need for entirely new reduction procedures for stabiliser arrays.
\section{Elementary Operations}
In this Section, we describe the allowed
elementary operations that transform a stabiliser array and which we will use to
reduce an array to its normal forms. As in the matrix case, these operations come in two kinds. The
first kind are the row operations. It is important to realise that row operations will not alter the
stabiliser state at all, but only alter the generator set it is represented by.
These are the \textit{row transposition},
which interchanges (transposes) two rows in the
stabiliser array, and the \textit{row multiplication}, which
multiplies one row with another one. The latter operation changes
the generators of the stabiliser group, but not the group itself
and hence not the stabiliser state either. We will use the phrase
``multiply row $k$ with row $l$'' to mean ``multiply rows $k$ and
$l$ elementwise and set row $l$ to the product obtained.'' The
multiplication table for Pauli operators is shown in Table
\ref{tab:mul}.
\begin{table}
$$
\begin{array}{r|rrrr}
\hline
&\id&X&Y&Z \\
\hline
\id&\id&X&Y&Z \\
X&X&\id&iZ&-iY \\
Y&Y&-iZ&\id&iX \\
Z&Z&iY&-iX&\id \\
\hline
\end{array}
$$
\caption{Multiplication table for Pauli operators; shown is $\sigma_{\text{row}}.\sigma_{\text{col}}$.
\label{tab:mul}}
\end{table}

The second kind of operations are the column operations, which may
alter the state. The column operations we will use are a certain class of
single-qubit operations, transposing two columns, and the CNOT
operation between two qubits. As single-qubit operations we take
those that act on one given column of the stabiliser operators by
permuting the Pauli operators (in the given column) among
themselves. These operations can be constructed from combinations
of Hadamard gates ($H$) and $\pi/4$ gates ($P$) (see Table
\ref{tab:1q}). Note that odd permutations must involve a
sign change in one of the Pauli operators in order to
correspond to a unitary operation. The particular sign
changes of Table \ref{tab:1q} have been chosen to make the
unitaries implementing the odd permutations involutory (apart from
a global phase). That is, $UU=\exp(i\phi)\id$. Note also that the
second and third permutation in the Table are each other's
inverse.
\begin{table}
$$
\begin{array}{rrrll}
\hline
X&Y&Z&& \mbox{Unitary}\\
\hline
X&Y&Z&&  \id \\
Z&X&Y&&  P H \\
Y&Z&X&&  H P^\dagger \\
-X&Z&Y&& P H P^\dagger \\
Y&X&-Z&& H P P H P^\dagger \\
Z&-Y&X&& H \\
\hline
\end{array}
$$
\caption{Truth table for the single-qubit operations employed by
the CNF algorithm. Any permutation of the set of Pauli operators
can be achieved. \label{tab:1q}}
\end{table}

For the bipartite normal form described in Section \ref{sec:bip}
we will need to divide the qubits into two parties
and only allow operations that are local to those parties.
Transposing two columns in the bipartite case is only allowed
when both columns (qubits) belong to the same party. Otherwise
this would be a non-local operation, which would very likely
affect the amount of entanglement.

The CNOT gate between two qubits, one being the control qubit and
one the target qubit, operates on the two corresponding columns of
the stabiliser array. In the case of a bipartite CNF, we must
again ensure that both qubits belong to the same party. The truth
table for the CNOT is given in table \ref{tab:cnot}. Note that the
column pertaining to the control qubit is modified too; this is a peculiarity
of the description of states by stabilisers.
\begin{table}
$$
\begin{array}{ll|rr}
\hline
C&T&C'&T'\\
\hline
\id & \id & \id & \id \\
\id & X & \id & X \\
\id & Y & Z&Y \\
\id & Z & Z&Z \\
X&\id & X&X \\
X&X&X&\id\\
X&Y&Y&Z\\
X&Z&-Y&Y\\
\hline
\end{array}
\qquad
\begin{array}{ll|rr}
\hline
C&T&C'&T'\\
\hline
Y&\id&Y&X\\
Y&X&Y&\id\\
Y&Y&-X&Z\\
Y&Z&X&Y\\
Z&\id&Z&\id\\
Z&X&Z&X\\
Z&Y&\id&Y\\
Z&Z&\id&Z\\
\hline
\end{array}
$$
\caption{Truth table for the CNOT gate employed by the CNF
algorithm. $C$ and $T$ refer to control and target qubit,
respectively. The primed columns give the values after the
operation. \label{tab:cnot}}
\end{table}
\section{Row-reduced Echelon Form}\label{sec:RREF}
While the Clifford Normal Form (CNF) of a stabiliser array will be
obtained below via application of both elementary row and column
operations, it is possible to obtain a normal form using elementary
row operations only. Due to its similarity to the matrix case, we
will call this normal form the \textit{Row-Reduced Echelon Form}
(RREF). The benefits of the RREF are that it is very easy to
obtain, the stabiliser state represented by the stabiliser array
is not changed, and it is applicable to states on any number of
parties. Furthermore, as we shall see below, it is an efficient
way to eliminate linearly dependent rows from the stabiliser
array.

The general structure of the RREF is most easily described in a recursive fashion.
There are three cases:
$$
\left(
\begin{array}{c|c}
\id &  \\
\vdots & \mbox{RREF}' \\
\id &
\end{array}
\right),\quad
\left(
\begin{array}{c|c}
\sigma &  * \ldots *  \\ \hline
\id    &              \\
\vdots & \mbox{RREF}' \\
\id    &
\end{array}
\right)
\mbox{ and }
\left(
\begin{array}{c|c}
\sigma_1 &  * \ldots * \\
\sigma_2 &  * \ldots * \\ \hline
\id    &               \\
\vdots &  \mbox{RREF}' \\
\id    &
\end{array}
\right).
$$
The symbols `$\vdots$' and `$\ldots$' denote a number of repeated
rows and columns. This number may be zero. The symbol
$\mbox{RREF}'$ denotes a (possibly empty) sub-array that is also
in RREF form. The symbol $*$ denotes either a Pauli operator or an
identity $\id$. Furthermore, $\sigma$, $\sigma_1$ and $\sigma_2$
are Pauli operators, and $\sigma_1$ and $\sigma_2$ anticommute. We
will refer to the operators in these positions as \textit{column
leaders} of their column, and \textit{row leaders} of their row.

The RREF algorithm works by applying a sequence of elementary row
operations to the stabiliser array. At every step of the algorithm
it is determined which elementary operation to apply based on the
values contained in a certain contiguous subarray of the full
array. At every step this subarray, which we will call the
\textit{active region}, either stays the same or decreases in
size. The algorithm terminates when the size of the active region
has decreased to zero. Note that the elementary operations operate
on the full stabiliser array and not just on the active region.

Let $K$, $N$ be the number of rows (generators) and columns
(qubits) of the stabiliser array, respectively. The variable $K_U$
contains the index of the first row in the active region,
and $N_L$ the index of its first column. The active region thus
consists of the array elements $(i,j)$ for $K_U\le i\le K$ and
$N_L\le j\le N$. Initially, the active region comprises the full
stabiliser array, hence $K_U=1$ and $N_L=1$.

In this and subsequent sections, the phase factors $\exp(i\phi_k)$
of the various generators will not be mentioned explicitly. They
are best maintained under the form of a single additional column
in the stabiliser array, which is modified by row permutations and
the elementary operations of tables \ref{tab:mul}, \ref{tab:1q}
and \ref{tab:cnot}.
\subsection{Algorithm RREF}
\begin{enumerate}
\item 
Count the number of different Pauli operators ($X$, $Y$ and $Z$) in the first column ($N_L$)
of the active region, i.e.\ restricting attention only to rows $K_U$ up to $K$.
\item 
Three cases can be considered:
    \begin{enumerate}
    \item 
    \textit{There are no Pauli operators in column $N_L$.}
        \begin{enumerate}
        \item Increase $N_L$ by 1.
        \end{enumerate}
    \item 
    \textit{There is only 1 kind of Pauli operator.}\\
    Let $k$ be the first row in the active region where column $N_L$ contains a Pauli operator.
        \begin{enumerate}
        \item Make row $k$ the top row of the active region by transposing, if necessary,
        row $k$ with row $K_U$.
        \item Multiply row $K_U$ with all other rows in the active region that have the same Pauli in
        column $N_L$.
        \item Increase $K_U$ and $N_L$ by 1.
        \end{enumerate}
    \item 
    \textit{There are at least 2 different kinds of Pauli operators.}\\
    Let $k_1$ be the first row in the active region where column $N_L$ contains a Pauli
    operator, and $k_2$ be the first row in the active region where column $N_L$ contains a
    different Pauli operator.
        \begin{enumerate}
        \item Make row $k_1$ the top row of the active region by transposing, if necessary,
        row $k_1$ with row $K_U$.
        \item Make row $k_2$ the second row of the active region by transposing, if necessary,
        row $k_2$ with row $K_U+1$.
        \item Multiply every other row in the active region with either row $K_U$, row $K_U+1$, both rows or none,
        depending on the element in column $N_L$ (see Table \ref{tab:xzc1}).
        \end{enumerate}
    \end{enumerate}
\item If the active region still has non-zero size ($N_L\le N$ and $K_U\le K$),
continue with step 1, else terminate.
\end{enumerate}
\begin{table}
\hrulefill \\
Initial stabiliser array:
$$
\left(
\begin{array}{c}
\sigma_1 \\
\sigma_2 \\
.
\end{array}
\right)
$$
Depending on the content of row 3, do the following:
\begin{itemize}
\item[$\id$:] Do nothing.
\item[$\sigma_1$:] Multiply row 1 with row 3.
\item[$\sigma_2$:] Multiply row 2 with row 3.
\item[$\sigma_3$:] Multiply row 1 with row 3, and then row 2 with row 3.
\end{itemize}
\hrulefill \caption{Required operations to eliminate any Pauli
operator from row 3 of the stabiliser array shown above. The
operators $\sigma_1$, $\sigma_2$, and $\sigma_3$ are a permutation
of $X$, $Y$ and $Z$. \label{tab:xzc1}}
\end{table}
\subsection{Checking independence of a set of generators}
The easiest way to check independence of a set of generators is to
compute the RREF of the stabiliser array. This fact is one other
property the stabiliser RREF and the matrix RREF have in common.
Dependencies between generators will show up as RREF rows
containing only $\id$ operators. Removing these all-$\id$ rows
leaves an independent set of generators.

\textit{Proof.} From the form of the RREF one observes that there
cannot be more than two rows with the same number of leading $\id$
operators, and if there are two such, they have a different row
leader. Consider a subset of generators $g_k$ in the RREF, having
$n_k$ leading $\id$ operators, and having $\sigma_k$ as row
leaders. Let the generators be sorted according to $n_k$ in
ascending order. When multiplying two rows that satisfy $n_2\ge
n_1$, the number of leading $\id$ operators in the product is
$n_1$ and the row leader is either $\sigma_1$ (if $n_2>n_1$) or
$\sigma_1\sigma_2$ (if $n_2=n_1$), which is different from either
$\sigma_1$ and $\sigma_2$. In both cases this shows that the
product cannot occur as another generator in the RREF. This proves
that it is not possible to write one RREF generator as a product
of other RREF generators. \qed
\subsection{Partial Trace of a Stabiliser State}\label{sec:ptrace}
A useful and important operation is the partial trace. The RREF algorithm is the central part
in the following efficient partial trace algorithm:

\medskip
\begin{center}
\textbf{\small Algorithm PTRACE}
\end{center}
\begin{enumerate}
\item By column permutations bring the columns of the
qubits to be traced out in first position.
\item Bring those columns to RREF.
\item Remove the rows containing the column leader(s).
\item Finally, remove those columns themselves.
\end{enumerate}

\medskip
\textit{Proof.}
To prove that this algorithm indeed calculates the partial trace,
consider again the three cases for the RREF:
$$
\left(
\begin{array}{c|c}
\id &  \\
\vdots & \mbox{RREF}' \\
\id &
\end{array}
\right),\quad
\left(
\begin{array}{c|c}
\sigma &  * \ldots *  \\ \hline
\id    &              \\
\vdots & \mbox{RREF}' \\
\id    &
\end{array}
\right)
\mbox{ and }
\left(
\begin{array}{c|c}
\sigma_1 &  * \ldots * \\
\sigma_2 &  * \ldots * \\ \hline
\id    &               \\
\vdots &  \mbox{RREF}' \\
\id    &
\end{array}
\right).
$$
We have to show that the state described by $\mbox{RREF}'$, say
$\rho'$, is the state obtained from the original stabiliser state
$\rho$ by tracing out the qubit pertaining to column 1. Denote the
sequences of $*$ operators by $g$, $g_1$ and $g_2$, respectively.

Using eq.~(\ref{eq:projmixed}), it is easy to see that, in the
first case,
$$
\rho = \frac{\identity}{2} \otimes \rho',
$$
and tracing out qubit 1 yields
$$
\trace_1\rho = \rho'.
$$
In the second case,
\beas
\rho &=& \frac{\identity\otimes\identity + \sigma\otimes g}{2} \,(\identity \otimes \rho') \\
&=& \frac{1}{2}(\identity \otimes \rho' + \sigma\otimes g\rho'),
\eeas
and again, as Pauli operators have trace 0,
$$
\trace_1\rho = \rho'.
$$
In the third and final case,
\beas
\rho &=& \frac{\identity\otimes\identity + \sigma_1\otimes g_1}{2} \,
\frac{\identity\otimes\identity + \sigma_2\otimes g_2}{2}\,
(\identity \otimes 2\rho') \\
&=& \frac{1}{2}(\identity \otimes \rho' + \sigma_1\otimes g_1\rho'+ \sigma_2\otimes g_2\rho'\\
&&+\sigma_1\sigma_2\otimes g_1 g_2\rho'),
\eeas
resulting yet again in
$$
\trace_1\rho = \rho'.
$$
\qed
\section{Single-Party Normal Form}
The CNF algorithm works by applying a sequence of elementary
operations to the stabiliser array. At every step of the algorithm
it is determined which elementary operation to apply based on the
values contained in a certain contiguous subarray of the full
array. At every step this subarray, which we will call the
\textit{active region}, either stays the same or decreases in
size. The algorithm terminates when the size of the active region
has decreased to zero. Note that the elementary operations operate
on the full stabiliser array and not just on the active region.

Let $K$, $N$ be the number of rows (generators) and columns
(qubits) of the stabiliser array, respectively. The variables
$K_U$ and $K_L$ contain the indices of the first
(uppermost) and last (lowermost) row in the active
region, and $N_L$ and $N_R$ the indices of its first
(leftmost) and last (rightmost) column. The active
region thus consists of the array elements $(i,j)$ for $K_U\le
i\le K_L$ and $N_L\le j\le N_R$. Initially, the active region
comprises the full stabiliser array, hence $K_U=1$, $K_L=K$,
$N_L=1$ and $N_R=N$. We will prove below that after every
iteration of the algorithm the stabiliser array has the block
structure
$$
\left(
\begin{array}{cccc|ccc|ccc}
X&\id&\ldots&\id  &  \id&\ldots&\id  & \id&\ldots&\id \\
\id&X&\ldots&\id  &  \id&\ldots&\id  & \id&\ldots&\id \\
\vdots&\vdots&\ddots&\vdots  &  \vdots&\vdots&\vdots &  \vdots&\vdots&\vdots      \\
\id&\id&\ldots&X  &  \id&\ldots&\id  & \id&\ldots&\id \\
\hline
\id&\id&\ldots&\id  &  *&\ldots&*    & \id&\ldots&\id \\
\vdots&\vdots&\vdots&\vdots  &  \vdots&\vdots&\vdots  &  \vdots&\vdots&\vdots      \\
\id&\id&\ldots&\id  &  *&\ldots&*    & \id&\ldots&\id
\end{array}
\right).
$$
The block containing the asterisks is the active region and has
not yet been brought to normal form. The columns on the left of
the active region correspond to qubits that are in an eigenstate
of the $X$ operator, the columns on its right correspond to qubits
that are in a totally mixed state. The final form, after
completion of the algorithm, is
$$
\left(
\begin{array}{cccc|ccc}
X&\id&\ldots&\id  &  \id&\ldots&\id \\
\id&X&\ldots&\id  &  \id&\ldots&\id \\
\vdots&\vdots&\ddots&\vdots  &  \vdots&\vdots&\vdots      \\
\id&\id&\ldots&X  &  \id&\ldots&\id \\
\hline
\id&\id&\ldots&\id  & \id&\ldots&\id \\
\vdots&\vdots&\vdots&\vdots  & \vdots&\vdots&\vdots      \\
\id&\id&\ldots&\id  & \id&\ldots&\id
\end{array}
\right).
$$
Here we have left open the possibility that the rows of the
initial stabiliser array might not be independent.
\subsection{Algorithm CNF1}
\begin{enumerate}
\item 
Count the number of different Pauli operators ($X$, $Y$ and $Z$)
in the first column ($N_L$) of the active region, i.e.\
restricting attention only to rows $K_U$ up to $K_L$.
\item 
Three cases can be considered:
    \begin{enumerate}
    \item 
    \textit{There are no Pauli operators in column $N_L$.}
        \begin{enumerate}
        \item If necessary, transpose column $N_L$ with column $N_R$.
        \item Decrease $N_R$ by 1.
        \end{enumerate}
    \item 
    \textit{There is only 1 kind of Pauli operator.}\\
    Let $k$ be the first row in the active region where column $N_L$ contains a Pauli operator.
        \begin{enumerate}
        \item Make row $k$ the top row of the active region by transposing, if necessary,
        row $k$ with row $K_U$.
        \item Apply whatever single-qubit operation on column $N_L$
        that brings that Pauli operator to an $X$.
        \item Multiply row $K_U$ with all other rows in the active region that have an $X$ in
        column $N_L$.
        \item Consider the elements of the first row of the active region (row $K_U$).
        To each of the columns beyond the first one that contains in the first row
        a Pauli different from $X$, apply a single-qubit operation to turn it into an $X$.
        \item To each of these columns, which now have an $X$ in the first row,
        successively apply a CNOT operation with control column $N_L$.
        \item Increase $K_U$ and $N_L$ by 1.
        \end{enumerate}
    \item 
    \textit{There are at least 2 different kinds of Pauli operators.}\\
    Let $k_1$ be the first row in the active region where column $N_L$ contains a Pauli
    operator, and $k_2$ be the first row in the active region where column $N_L$ contains a
    different Pauli operator.
        \begin{enumerate}
        \item Make row $k_1$ the top row of the active region by transposing, if necessary,
        row $k_1$ with row $K_U$.
        \item Make row $k_2$ the second row of the active region by transposing, if necessary,
        row $k_2$ with row $K_U+1$.
        \item Bring the element on row $K_U$ to an $X$ and the element on row $K_U+1$ to a $Z$
        by applying, if necessary, a single-qubit operation on column $N_L$.
        \item Consider the first two rows of the active region (rows $K_U$ and $K_U+1$).
        Find the first column beyond column $N_L$, say column $l$,
        that contains an anticommuting pair on those rows
        (i.e.\ two non-identical Pauli operators).
        \item Bring the anticommuting pair to an $(X,Y)$ pair by applying, if necessary,
        a single-qubit operation to that column.
        \item Apply a CNOT operation to that column, with column $N_L$ as control.
        \item The extent of the active region is not changed in this case.
        \end{enumerate}
    \end{enumerate}
\item If the active region still has non-zero size ($N_L\le N_R$ and $K_U\le K_L$),
continue with step 1, else terminate.
\end{enumerate}
\subsection{Proof of correctness of algorithm CNF1}
We will now show that algorithm CNF1 indeed brings any stabiliser
array into its normal form. We consider the three cases (a), (b)
and (c) in succession.
\subsubsection{Case (a)}
This case corresponds to column $N_L$ containing $\id$ only and
therefore belongs to the block right of the active region. Step
(a.i) does just that and step (a.ii) subsequently excludes this
column from the active region.
\subsubsection{Case (b)}
This case corresponds to a column containing $\id$ operators and
Pauli operators of just one kind. Step (b.i) brings the first of
these Pauli operators to the top row, with the ultimate goal of
excluding this row from the active region. Step (b.ii) applies a
single-qubit rotation to bring the Pauli operators in standard
form, which in this case is an X operator.

In step (b.iii) the column is then ``cleaned up''. Through
multiplying the top row $K_U$ with other rows containing an $X$ in
column $N_L$, we obtain a stabiliser array that is still
describing the same state but contains only one $X$ in column
$N_L$. So this column is already in standard form.

However, the top row is not in standard form yet. Step (b.iv)
applies an appropriate single-qubit operation to every column in
the active region, except for the first one, so that the first row
contains either $\id$ or $X$ operators. Step (b.v) then performs a
``row cleanup'', by applying CNOTs to the columns starting with an
$X$, the first column being the control column. The target $X$
operators are thereby turned into $\id$, leaving us with a first
row of the form $(X,\id,\ldots,\id)$.

It is not a priori clear, however, that step (b.v) is not undoing
the cleanup of column $N_L$ by step (b.iii). Nevertheless,
inspection of the CNOT truth table reveals that the $\id$
operators in column $N_L$ can either be turned into a $Z$ or
remain $\id$, by any number of CNOTs. Although a $Z$ operator can
actually occur during the execution of step (b.iii), in the end
all operators will be turned back into $\id$. This must be so
because the top row of the active region is turned into
$(X,\id,\ldots,\id)$, which does not commute with a row starting
with a $Z$. So the assumption of commutativity of the generators
ensures that step (b.iii) is not undone by step (b.v).

Finally, we note that both the first row and the first column are
now in standard form and can be removed from the active region,
which is done in step (b.vi). The top left block in the normal
form array hereby receives one further $X$ operator.
\subsubsection{Case (c)}
The most difficult case to investigate is case (c), because here
it is not a priori clear that any progress is made within an
iteration. Indeed, the extent of the active region is not changed
and it is not clear that further iterations will eventually escape
from case (c), resulting potentially in an infinite loop.

However, every execution of case (c) does result in measurable
progress. As can be seen from the truth table of the CNOT
operation, the end result of the CNOT in step (c.vi) is
$$
\begin{array}{r|ll|ll}
&N_L&l&N_L'&l' \\
\hline
K_U&X&X&X&\id \\
K_U+1&Z&Y&\id&Y \\
\hline
\end{array}
$$
Hence, a $\id$ is introduced in row $K_U$ where there originally
was none. Furthermore, no further algorithmic step in case (c)
ever touches this element again, by the following reasoning.
\begin{itemize}
\item The only operations that do change the top row $K_U$ are the transposition in step (c.i),
and the CNOT in step (c.vi).
\item
Step (c.i) is executed at most once before the algorithm breaks
out of the (c) case, namely at the very beginning. This is because
the $X$ brought in the top left position is not changed by the
CNOT.
\item
The first CNOT that operates on target column $l$ introduces the
$\id$ there. In further iterations, the CNOT will not operate on
column $l$ a second time, because step (c.iv) sets the target
column to a column containing an anticommuting pair in the top two
rows, and the $\id$ created in the top of column $l$ does not form
part of an anticommuting pair.
\end{itemize}
It is now easy to see why the algorithm must eventually break out
of the (c) case. Every iteration through this case increases the
number of $\id$ operators in the top row by 1, but there are only
a limited number of places (columns) available to do this. Hence
the number of successive iterations through case (c) must be
limited too.

The algorithm breaks out of the loop through the (c) case when
there are no further anticommuting pairs in column $N_L$. As a
consequence, the algorithm will then either execute case (a) or
case (b), thereby again reducing the extent of the active region.
\subsection{Alternative proof of projection formulas (\ref{eq:projpure}) and (\ref{eq:projmixed})}
In this subsection we present a new proof of the equivalence of the expressions
(\ref{eq:pureorig}) and (\ref{eq:projpure})
for a pure stabiliser state, and of
(\ref{eq:mixedorig}) and (\ref{eq:projmixed})
for a mixed stabiliser state.

By the proof of the CNF1 algorithm, a state described by a certain
stabiliser array is unitarily equivalent to the state described by
the normal form of that array. Let the initial stabiliser group
$S$ be given by a stabiliser array. Let $S'$ be the stabiliser
group described by the normal form of that array. The $K$
generators of $S'$ are of the form
$$
g'_k = \id\otimes\ldots\otimes X\otimes\ldots\otimes\id,
$$
with the $X$ operator in the $k$-th tensor factor.
The stabiliser state corresponding to the normal form
is therefore
\beas
\rho' &=& \frac{1}{2^N} \sum_{i_1,\ldots,i_K\in\{0,1\}} X^{i_1}\otimes \ldots \otimes X^{i_K}
\otimes\id^{\otimes N-K} \\
&=& ((\id+X)/2)^{\otimes K} \otimes (\id/2)^{\otimes N-K} \\
&=& \frac{1}{2^{N-K}} \prod_{k=1}^K \frac{\id+g'_k}{2}. \eeas Let
$U$ be the unitary corresponding to the sequence of elementary
operations that brought the stabiliser array to its normal form.
To wit, $S$ consists of the elements $g=Ug'U^\dagger$, $g'\in S'$,
and can be generated by the generators $g_k:=Ug'_k U^\dagger$.
Then the stabiliser state corresponding to $S$ is given by \beas
\rho &=& \frac{1}{2^N} \sum_{g\in S} g \\
&=& \frac{1}{2^N} \sum_{g'\in S'} Ug'U^\dagger \\
&=& U\rho' U^\dagger \\
&=& \frac{1}{2^{N-K}} \prod_{k=1}^K U\frac{\id+g'_k}{2}U^\dagger \\
&=& \frac{1}{2^{N-K}} \prod_{k=1}^K \frac{\id+g_k}{2}.
\eeas
\qed
\section{Fidelity between stabiliser states}
The topic of this section is an algorithm to calculate
the overlap $F=\trace[\rho_1 \rho_2]$ between two mixed stabiliser
states $\rho_1$ and $\rho_2$ directly from their $K_1\times N$ and
$K_2\times N$ stabiliser arrays $A_1$ and $A_2$.

While the overlap between two states is certainly an interesting quantity, the Bures distance
$$
D(\rho_1,\rho_2):=2\sqrt{1-F_u(\rho_1,\rho_2)},
$$
where $F_u$ is the Uhlmann fidelity
$$
F_u(\rho_1,\rho_2):=\trace[\sqrt{\sqrt{\rho_1}\rho_2\sqrt{\rho_1}}],
$$
is a much more desirable quantity, as it is an actual distance
measure and has a much nicer interpretation. It is well-known that
for pure states the Uhlmann fidelity between two states is just
the square root of their overlap, while for general mixed states
there is no such relation.
It will turn out that with just a minor modification the algorithm is also
able to calculate the Uhlmann fidelity.
This allows us to calculate the overlap, the Uhlmann
fidelity and Bures distance for stabiliser states in one go.

For the calculation of the overlap (and fidelity) it is imperative to take the
generator \textit{phases} into account. We will use the vectors
$S_1$ and $S_2$ for that purpose. The elementary row operations of
multiplication and permutation of stabiliser rows are understood
to treat the phase vector as an additional column of the
stabiliser array. Furthermore, row multiplication, single-qubit
rotation and CNOT operation have to multiply the appropriate
generator phase with the phase factor mentioned in their truth
tables.
\subsection{Algorithm OVERLAP}
\begin{enumerate}
\item 
Construct the $(K_1+K_2)\times N$ composite array $A$ and the composite vector $S$ of generator phases:
$$
A = \left(
\begin{array}{c}
A_1 \\ \hline
A_2
\end{array}
\right)
\quad
S = \left(
\begin{array}{c}
S_1 \\ \hline
S_2
\end{array}
\right).
$$
\footnote{This composite array is no longer a stabiliser array
because generators from $A_1$ need not commute with those of
$A_2$. Even worse, $A_1$ and $A_2$ may generate conflicting stabilisers, i.e.\
with opposite phase factors.
One thus cannot just apply reduction algorithms, say RREF, to $A$ as a whole,
and claim to have calculated the rank of $A$; in fact, the rank of $A$ is just not defined.
There are, however, some situations where one can treat the composite array, with due care,
as if it were a stabiliser array.
This is especially useful when one needs to apply the same column operations to
$A_1$ and $A_2$.}
\item 
By applying the CNF1 algorithm to the composite array $A$ (and its
vector $S$ of generator phases), with initial active region set to
the full $A_1$ part (excluding $A_2$!), the $A_1$ part is brought to CNF form, while
automatically applying the same sequence of column operations to
the $A_2$ part. Let $R_1$ be the number of $X$ operators in this
CNF.
\item 
Set the active region to all the rows of the $A_2$ part and all
the columns for which $A_1$ contains $X$ operators. That is,
$K_U=K_1+1$, $K_L=K_1+K_2$, $N_L=1$ and $N_R=R$. Set $T=1$.
\item 
Count the number of different Pauli operators ($X$, $Y$ and $Z$)
in the first column ($N_L$) of the active region, i.e.\
restricting attention only to rows $K_U$ up to $K_L$.
\item 
Three cases can be considered:
    \begin{enumerate}
    \item
    \textit{There are no Pauli operators in column $N_L$.}
        \begin{enumerate}
        \item Do nothing.
        \end{enumerate}
    \item
    \textit{There is only 1 kind of Pauli operator.}\\
    Let $k$ be the first row in the active region where column $N_L$ contains a Pauli operator.
        \begin{enumerate}
        \item Make row $k$ the top row of the active region by transposing, if necessary,
        row $k$ with row $K_U$.
        \item Multiply row $K_U$ with all other rows in the active region that have a Pauli operator
        (necessarily equal to the one on row $K_U$) in column $N_L$.
        \item Let $P$ be the element in column $N_L$ on row $K_U$ (the column leader).
        \begin{itemize}
            \item If $P$ is not an $X$, divide $T$ by 2 and increase $K_U$ by 1.
            \item If $P$ is an $X$, multiply row $N_L$ (that is, the row containing an $X$ in
            column $N_L$ of subarray $A_1$) to row $K_U$.
        \end{itemize}
        \end{enumerate}
    \item
    \textit{There are at least 2 different kinds of Pauli operators.}\\
    Let $k_1$ be the first row in the active region where column $N_L$ contains a Pauli
    operator, and $k_2$ be the first row in the active region where column $N_L$ contains a
    different Pauli operator.
        \begin{enumerate}
        \item Make row $k_1$ the top row of the active region by transposing, if necessary,
        row $k_1$ with row $K_U$.
        \item Make row $k_2$ the second row of the active region by transposing, if necessary,
        row $k_2$ with row $K_U+1$.
        \item Multiply every other row in the active region with either row $K_U$, row $K_U+1$, both rows or none,
        depending on the element in column $N_L$ (according to Table \ref{tab:xzc1}).
        \item Let $P_1$ be the element in column $N_L$ on row $K_U$, and $P_2$ the one on row $K_U+1$
        (the column leaders). Turn $P_2$ into an $X$, as follows:
        \begin{itemize}
            \item If neither $P_1$ nor $P_2$ is an $X$, multiply row $K_U$ to row $K_U+1$, effectively turning $P_2$
            into an $X$.
            \item If $P_1$ is an $X$, transpose row $K_U$ with $K_U+1$.
        \end{itemize}
        \item Multiply row $N_L$ (that is, the row containing an $X$ in column $N_L$ of subarray $A_1$) to
        row $K_U+1$.
        \item Divide $T$ by 2, and increase $K_U$ by 1.
        \end{enumerate}
    \end{enumerate}
\item Increase $N_L$ by 1.
\item If the active region is still non-empty
($N_L\le N_R$ and $K_U\le K_R$), continue with step 4.
\item (End Game)
Here we calculate a correction factor $C$ for the overlap and the fidelity.
Set $C=1$ as default value.
If $K_U\le K_L$ do the following.
    \begin{enumerate}
    \item Case $N_L\le N$:
        Consider the bottom right block of stabiliser array $A_2$ consisting of rows
        $K_U$ to $K_L$ and columns $N_R+1$ to $N$.
        Calculate the rank $R_2$ of that block using, e.g.\ the RREF algorithm.
        From $R_2$ calculate the correction factor as $C=2^{N-R_1-R_2}$.
    \item Case $N_L>N$:
        Let $t_k$ be the generator phase of row $k$. If at least one of the $t_k$ for $K_U\le k\le K_L$
        is $-1$, set $T=0$.
    \end{enumerate}
\item Terminate with return values $F=CT/2^{N-K_1+N-K_2}$ for the overlap
and $F_u = C\sqrt{T/2^{N-K_1+N-K_2}}$ for the Uhlmann fidelity.
\end{enumerate}
\subsection{Proof of correctness of algorithm OVERLAP}
The overlap $F=\trace[\rho_1 \rho_2]$ can be calculated
iteratively by performing the trace as a succession of partial
traces over single qubits: $F=\trace[\trace_1[\rho_1 \rho_2]]$,
where $\trace_1$ denotes the partial trace over the first qubit.
What we need to show is that one iteration of steps 4-7 indeed
performs this single-qubit partial trace. It will be convenient to
express the overlap in terms of the projectors $\cP_1$ and
$\cP_2$, with $\rho_1 = \cP_1/2^{N-K_1}$ and $\rho_2 =
\cP_2/2^{N-K_2}$. Then
$$
F = \frac{1}{2^{N-K_1+N-K_2}}\trace[\cP_1 \cP_2].
$$
Keeping in mind that we also want to calculate $F_u$,
we will proceed by first calculating $\cP_1 \cP_2 \cP_1$.
The overlap is just the trace of this quantity, which is the same as
$\trace[\cP_1 \cP_2]$ by virtue of $\cP_1$ being a projector.

Step 2 of the algorithm applies the same sequence of unitaries to
both states, hence the overlap between them does not change (and
neither does the Uhlmann fidelity). Let $\cP_1$ thus be specified
by a CNF stabiliser array, containing $R_1\le N$ $X$-operators:
$$
\cP_1 = \bigotimes_{i=1}^{R_1} \frac{\id+s_i X}{2} \otimes \id^{\otimes N-R_1}.
$$
If all generators in array $A_1$ are independent, we obviously have $R_1=K_1$.
In the above expression, $s_i$ is the generator phase of the $i$-th generator of
$\cP_1$. Likewise, we will denote by $t_i$ the generator phase of
the $i$-th generator of $\cP_2$. Furthermore, let $\cP_1'$ be the
stabiliser projector of the array obtained by deleting row 1 and
column 1 from $A_1$.

In the following, we will calculate $\cP_1 \cP_2 \cP_1$ and show that it is
equal to a certain scalar value $T$ times a tensor product of rank-1 projectors
and identity operators. We will proceed in an iterative fashion, by showing that
$\cP_1 \cP_2 \cP_1$ decomposes as a scalar $T_1$ times either a rank-1 projector or an identity
tensored with a product $\cP'_1 \cP'_2 \cP'_1$ of projectors over qubits 2 to $N$.
To calculate $T$, we start off with the initial value $T=1$
and update $T$ by multiplying it with the value of $T_1$ found at each iteration.

We will assume first that $A_1$ and $A_2$ have more than 1 column. The
case that they only have 1 column, which is what can happen in the
final iteration of the algorithm, will be considered in subsection
4.
We will also assume that the first tensor factor of $\cP_1$ is a rank-1 projector (i.e.\ $R_1>0$).
The case that $\cP_1$ equals the identity (which will again typically happen at the end of the iterations)
will also be covered in subsection 4.

Let us now take on the main case, where there are at least two tensor factors to consider, and the
first factor of $\cP_1$ is $\frac{\id+s_i X}{2}$. Thus we can write
$\cP_1=\frac{\id+s_i X}{2}\otimes \cP'_1$.
As in algorithm CNF1 there are three cases to consider,
depending on the number of different Pauli operators contained in
the first column of the second array $A_2$. We will investigate these
three possibilities in succession.

It is useful to note that
$$
\frac{\id+sX}{2}\,\sigma\,\frac{\id+sX}{2} =
\left\{
\begin{array}{ll}
(\id+sX)/2, & \sigma=\id \\ [2mm]
s(\id+sX)/2, & \sigma=X \\ [2mm]
0, & \sigma=Y,Z
\end{array}
\right.
$$

\subsubsection{Case (a)}
If the first column of $A_2$ contains no Pauli operators, this
corresponds to $\cP_2$ being of the form
$$
\cP_2 = \id\otimes \cP_2',
$$
where $\cP_2'$ is the stabiliser projector of the array obtained
by deleting column 1 from $A_2$. Hence
\beas
\cP_1 \cP_2 \cP_1
&=& (\frac{\id+s_1 X}{2}\otimes \cP_1') \,(\id\otimes\cP_2')\,(\frac{\id+s_1 X}{2}\otimes \cP_1') \\
&=& \frac{\id+s_1 X}{2} \otimes \cP_1'\,\cP_2'\,\cP_1'.
\eeas
This is indeed of the form claimed above, with scalar value $T_1=1$.
Hence, nothing needs to be done in this iteration except for
deleting column 1.
\subsubsection{Case (b)}
Steps (b.i) and (b.ii) bring column 1 of $\cP_2$ to RREF form. In this case, column 1 will contain a single Pauli
operator, $\sigma$, in row 1. Denote the remaining operators on row 1 by $g'$. Let $\cP_2'$ be the
stabiliser projector of the array obtained by deleting row 1 and column 1 from $A_2$.
Thus $\cP_2$ is of the form
$$
\cP_2 = \frac{\id+t_1\sigma\otimes g'}{2}(\id\otimes\cP_2').
$$
We then have
\beas
\cP_1\, \cP_2\,\cP_1
&=& (\frac{\id+s_1 X}{2}\otimes \cP_1') \,\frac{\id+t_1\sigma\otimes g'}{2}\\
&& \times\,(\id\otimes\cP_2')\,(\frac{\id+s_1 X}{2}\otimes \cP_1') \\
&=& \frac{1}{2}\frac{\id+s_1X}{2}\otimes\cP_1' \cP_2' \cP_1' \\
&& +\,\frac{t_1}{2} \,\frac{\id+s_1X}{2}\,\sigma\,\frac{\id+s_1X}{2} \otimes \cP_1' g' \cP_2' \cP_1'.
\eeas
We can therefore distinguish two cases.
If $\sigma$ is not an $X$, we find
$$
\cP_1\, \cP_2\,\cP_1 = \frac{1}{2}\frac{\id+s_1X}{2}\otimes\cP_1' \cP_2' \cP_1'.
$$
This corresponds to a value of $T_1=1/2$.
This is implemented in step (b.iii, first case) by dividing the
running $T$ by 2, and deleting row 1 and column 1 from $A_2$.

If, on the other hand, $\sigma=X$, we have
\beas
\cP_1\, \cP_2\,\cP_1 &=& \frac{1}{2}\frac{\id+s_1X}{2}\otimes\cP_1' \cP_2' \cP_1' \\
&&+ \,\frac{s_1t_1}{2} \,\frac{\id+s_1X}{2}\otimes \cP_1' g' \cP_2' \cP_1' \\
&=& \frac{\id+s_1X}{2} \otimes \cP_1'\,\frac{\id+s_1t_1g'}{2}\,\cP_2'\,\cP_1' \\
&=& \frac{\id+s_1X}{2} \otimes \cP_1'\,\cP_2''\,\cP_1'.
\eeas
where $\cP_2''=\frac{\id+s_1t_1g'}{2}\,\cP_2'$ is a projector.
This corresponds to a scalar value of $T_1 = 1$.
This is accomplished in step (b.iii, second case) by multiplying
row 1 of $A_1$ to row 1 of $A_2$, and deleting column 1 of $A_2$
(leaving row 1).
\subsubsection{Case (c)}
Steps (c.i), (c.ii) and (c.iii) bring column 1 of $A_2$ in RREF
form. Column 1 will contain two Pauli operators, $\sigma_1$ in row
1, and $\sigma_2\neq\sigma_1$ in row 2. Step (c.iv) ensures, by
suitable row multiplication or transposition, that $\sigma_2$ is
an $X$ operator, so $\sigma_1$ is not. Let the remaining operators
on rows 1 and 2 be denoted by $g_1'$ and $g_2'$, respectively. Let
$\cP_2'$ be the stabiliser projector of the array $A_2'$, obtained
by deleting rows 1 and 2 and column 1 from $A_2$. Then $\cP_2$ is
given by
$$
\cP_2=\frac{\id+t_1\sigma_1\otimes g_1'}{2}\,\frac{\id+t_2X\otimes g_2'}{2}\,(\id\otimes \cP_2').
$$
Thus
\beas
\lefteqn{\cP_1 \,\cP_2\,\cP_1} \\
&=& (\frac{\id+s_1 X}{2}\otimes \cP_1') \\
 && \times\, \frac{\id+t_1\sigma_1\otimes g_1'}{2}\,\frac{\id+t_2X\otimes g_2'}{2} \\
 && \times\, (\id\otimes \cP_2')\,(\frac{\id+s_1 X}{2}\otimes \cP_1') \\
&=& \frac{1}{4}\Big[
 \frac{\id+s_1 X}{2} \otimes \cP_1'\cP_2'\cP_1' \\
&& \quad +\,t_1 \frac{\id+s_1 X}{2}\sigma_1\frac{\id+s_1 X}{2}\otimes \cP_1' g_1' \cP_2' \cP_1' \\
&& \quad +\,t_2 \frac{\id+s_1 X}{2} X \frac{\id+s_1 X}{2} \otimes \cP_1' g_2' \cP_2' \cP_1' \\
&& \quad +\,t_1 t_2 \frac{\id+s_1 X}{2} \sigma_1 X \frac{\id+s_1 X}{2} \otimes \cP_1' g_1' g_2' \cP_2' \cP_1'
\Big] \\
&=&  \frac{1}{4}\Big[
 \frac{\id+s_1 X}{2} \otimes \cP_1'\cP_2'\cP_1' \\
&& \quad +\,s_1 t_2 \frac{\id+s_1 X}{2} \otimes \cP_1' g_2' \cP_2' \cP_1'
\Big],
\eeas
giving
\beas
\cP_1 \,\cP_2\,\cP_1 &=&
\frac{1}{2} \frac{\id+s_1 X}{2} \otimes \cP_1' \frac{\id+s_1 t_2 g_2'}{2} \cP_2'\cP_1' \\
&=& \frac{1}{2} \frac{\id+s_1 X}{2} \otimes \cP_1' \cP_2''\cP_1'
\eeas
with $\cP_2'' = \frac{\id+s_1 t_2 g_2'}{2} \cP_2'$ a projector.
This corresponds to $T_1=1/2$.
This is implemented in steps (c.v) and (c.vi) through multiplying
row 2 in $A_2$ by row 1 of $A_1$, dividing $T$ by 2, and
subsequently deleting row 1 and column 1 in $A_2$.
\subsubsection{End Game}
We still have to consider the situation where
there is only one column left and the one where $\cP_1$ is a tensor product of identity operators.

The first situation is when $N_L=N$. In that case the symbols $\cP_1'$, $g'$ and $\cP_2'$ used in the
previous subsections are meaningless. However, we can still make
sense out of the calculations if we replace these symbols formally
by the scalar 1. Inspection of the relevant calculations then
shows that at the very end of the algorithm, if $N_L=N$ we have to
check whether one of the remaining generator phases $t_i$ is $-1$,
in which case the states under consideration are orthogonal.
That means both the overlap and the Uhlmann fidelity are 0, which we impose by
setting $T=0$.

If $N_L<N$ but $\cP_1$ acts as the identity on columns $N_L$ to $N$,
$\cP_1 \cP_2 \cP_1$ reduces to $\cP_2$. This can easily be decomposed as a tensor product by calculating
its rank $R_2$ (the easiest way to do this is by using the RREF algorithm).
Thus the remaining part on columns $N_L$ to $N$ of $\cP_1 \cP_2 \cP_1 = \cP_2$ is unitarily equivalent to
$$
\bigotimes_{i=1}^{R_2} \frac{\id+X}{2} \otimes \id^{\otimes N-R_1-R_2}.
$$
\subsubsection{Overlap and Uhlmann Fidelity}
In the previous subsection we have shown that
$\cP_1 \cP_2 \cP_1$ is equal to $T$ times a tensor product of $R_1+R_2$ rank-1 projectors
and $N-R_1-R_2$ identity operators.
Calculating the overlap and the Uhlmann fidelity is now easy.
Assuming that $R_1=K_1$, we have for the overlap
\bea
F &=& \frac{1}{2^{N-K_1+N-K_2}} \trace[\cP_1 \cP_2 \cP_1] \nonumber \\
&=& \frac{1}{2^{N-K_1+N-K_2}}\, T\, 2^{N-R_1-R_2} \nonumber \\
&=& 2^{-(N-K_2+R_2)}\, T,
\eea
where it has to be noted that $R_2\le K_2$.
Since $T$ is also a negative power of $2$ one sees that the overlap takes values of either 0 or $2^{-j}$,
where $j$ is an integer between 0 and $N$.

Similarly, the Uhlmann fidelity between states $\rho_1$ and $\rho_2$ is given by
$$
F_u = \trace[\sqrt{\sqrt{\rho_1}\rho_2\sqrt{\rho_1}}].
$$
Again we substitute the stabiliser states for their appropriately
scaled projectors. Noting that the square root of a projector is
that same projector gives
\bea
F_u &=& (2^{N-K_1}\,2^{N-K_2})^{-1/2} \,\trace[\sqrt{\cP_1 \cP_2 \cP_1}] \nonumber \\
&=& (2^{N-K_1}\,2^{N-K_2})^{-1/2} \, \sqrt{T}\, 2^{N-R_1-R_2} \nonumber \\
&=& 2^{(K_2-K_1)/2-R_2}\,\sqrt{T}.
\eea
\section{Bipartite Normal Form\label{sec:bip}}
In this Section, we will modify the single-party algorithm CNF1 so
that it can be used to reduce a stabiliser array of a bipartite
system to a certain normal form. This algorithm will allow us to
deduce the exact structure of this normal form, which is the
content of Theorem \ref{th:cnf}. This Theorem basically tells us
that a bipartite mixed stabiliser state is locally equivalent to a
tensor product of some number of pure EPR pairs and a separable
mixed state. The main benefit of this normal form is that the
entanglement of the state can immediately be read off from the
normal form. Because of the Theorem, it turns out that in order to
calculate the state's entanglement it is not necessary to actually
compute the normal form completely. Instead, a simplified
algorithm to calculate entanglement will be presented.

Let us start with the statement of the normal form.
\begin{theorem}\label{th:cnf}
Consider a system of $N$ qubits, separated into two parties, A and
B, containing $N_A$ and $N_B$ qubits, respectively. Consider a
stabiliser state described by an array of $K$ independent,
commuting generators.

i) By applying a suitable sequence of elementary row operations
and local elementary column (qubit) operations, the stabiliser
array can be brought into the following normal form:
\be\label{eq:cnf2} \left(
\begin{array}{cccc|ccc||cccc|ccc}
X&\id&\ldots&\id  &  \id&\ldots&\id  &X&\id&\ldots&\id  &  \id&\ldots&\id \\
Z&\id&\ldots&\id  &  \id&\ldots&\id  &Z&\id&\ldots&\id  &  \id&\ldots&\id \\
\id&X&\ldots&\id  &  \id&\ldots&\id  &\id&X&\ldots&\id  &  \id&\ldots&\id \\
\id&Z&\ldots&\id  &  \id&\ldots&\id  &\id&Z&\ldots&\id  &  \id&\ldots&\id \\
\vdots&\vdots&&\vdots &  \vdots&&\vdots &\vdots&\vdots&&\vdots &  \vdots&&\vdots \\
\id&\id&\ldots&X  &  \id&\ldots&\id  &\id&\id&\ldots&X  &  \id&\ldots&\id \\
\id&\id&\ldots&Z  &  \id&\ldots&\id  &\id&\id&\ldots&Z  &  \id&\ldots&\id \\
\hline
\id&\id&\ldots&\id & *&\ldots&* &\id&\id&\ldots&\id & *&\ldots&* \\
\vdots&\vdots&&\vdots & *&\ldots&* &\vdots&\vdots&&\vdots & *&\ldots&* \\
\id&\id&\ldots&\id & *&\ldots&* &\id&\id&\ldots&\id & *&\ldots&*
\end{array}
\right)
\ee
Here, the asterisk stands for either $\id$ or $X$,
and the double line is the separation line between the two parties.

ii) Every pair of rows containing the $XZ$ combinations
corresponds to two qubits (one from each party) being in a pure
maximally entangled EPR state and completely disentangled from the
other qubits. The rows in the lower blocks of the normal form,
containing only $\id$ and $X$ operators, correspond to the
remaining qubits being in a (in general, mixed) separable state.

iii) The stabiliser state described by the stabiliser array is
locally equivalent to a tensor product of a certain number $p$ of
EPR pairs $\Psi$ with a separable state. For any additive
entanglement measure $E$, the entanglement of the stabiliser state
is $p E(\Psi)$. An upper bound on $p$ is given by \be p \le
\min(\lfloor K/2\rfloor, N_A, N_B). \ee Equality is obtained if
and only if $K = 2 N_A = 2 N_B$.
\end{theorem}

We will postpone the proof of part i) of the Theorem, the normal
form, to the end of this Section. The proof of part iii), the
entanglement properties of the normal form, is elementary and is
left to the reader. The proof of part ii) is presented next.

\textit{Proof of part ii).} For convenience of notation, we first
permute the qubits in such a way that the pairs of columns having
an $XZ$ pair in the same rows are adjacent. By
eq.~(\ref{eq:projmixed}), the stabiliser state corresponding to
the normal form of the Theorem is
$$
\rho = \frac{1}{2^{N-K}} \prod_{k=1}^{2p} \frac{\id+g_k}{2}
\prod_{l=2p+1}^K \frac{\id+g_l}{2}.
$$
From the specific form of the generators one sees that $\rho$ can
be written as tensor product
$$
\rho= \left(\frac{\id+X\otimes X}{2}\,\frac{\id+Z\otimes Z}{2}\right)^{\otimes p}
\otimes\rho',
$$
where $\rho'$ corresponds to the factor $\prod_{l=2p+1}^K
(\id+g_l)/2$ containing $\id$ and $X$ operators only. It is a
simple matter to verify that the factor $(\id+X\otimes
X)(\id+Z\otimes Z)/4$ is identical to the EPR state $\Psi =
\ket{\psi}\bra{\psi}$, with $\psi=(1,0,0,1)^T/\sqrt{2}$.

It is also simple to see that $\rho'$ is a separable state. As it
only contains $\id$ and $X$ operators, it is diagonal in any basis
where $X$ is diagonal, and it is well-known and easy to see that
diagonal states are separable. \qed
\subsection{Algorithm CNFP for calculating the entanglement}
We now present an algorithm to calculate the number of EPR pairs
in the normal form, without actually reducing the stabiliser array
completely to that normal form. This algorithm is almost identical
to algorithm CNF1, the reduction algorithm for the single-party
case.

To calculate the entanglement, the initial active region is set to
comprise the block of elements pertaining to party A only, rather
than the full stabiliser array, and algorithm CNFP (CNF for a
single Party) is run on this active region.

Algorithm CNFP is identical to CNF1, apart from the following two
differences:

\textit{Step (b.vi)}. While in CNF1 step (b.iii) is never undone
by step (b.v) due to commutativity of the generators, this need no
longer be the case here. Indeed, here we restrict attention to one
party only, and the parts of the generators local to party A need
not commute. Hence step (b.v) might leave $Z$ operators in the
leftmost column of the active region. We thus need a modification
here: first we must check whether this has happened and only if
there are no $Z$ operators in this column may $K_U$ and $N_L$ be
increased by 1. Otherwise, the extent of the active region must
stay the same. The additional $Z$'s will then be treated in the
next iteration of the algorithm.

\textit{Step (c.iv)}. In step (c.iv), the original algorithm
looked for an anticommuting pair in the top two rows, the presence
of which having been guaranteed by generator commutativity. Here,
again, this is no longer true, because the pair might be located
in party B, which we are not allowed to touch here. We therefore
need a second modification, to deal with the case that there is no
such anticommuting pair. In that case, instead of steps (c.v),
(c.vi) and (c.vii), the following operations must be executed.
Recall that the first column has an $XZ$ pair in its first two
rows. This pair can now be used to eliminate all other Pauli
operators in both the first two rows (by suitable single-qubit
operations and CNOTs) and in the first column (by suitable row
multiplications). Tables \ref{tab:xzr} and \ref{tab:xzc} contain
the details. After that, the $XZ$ pair can be split off from the
active region to form part of the normal form, by increasing $K_U$
by 2, and $N_L$ by 1.
\begin{table}
\hrulefill \\
Initial stabiliser array:
$$
\left(
\begin{array}{cc}
X&. \\
Z&.
\end{array}
\right)
$$
Depending on the content of column 2, do the following:
\begin{itemize}
\item[$\id X$:]
\item[$\id Y$:]
\item[$\id Z$:]
Using a single-qubit operation, bring column 2 to $\id Z$, then
perform a CNOT with column 1 as target (!) and column 2 as
control.
\item[$X \id$:]
\item[$Y \id$:]
\item[$Z \id$:]
Using a single-qubit operation, bring column 2 to $X \id$, then
perform a CNOT with column 1 as control and column 2 as target.
\item[$XX$:]
\item[$YY$:]
\item[$ZZ$:]
Using a single-qubit operation, bring column 2 to $ZZ$, then
perform a CNOT with column 1 as target (!) and column 2 as
control. Column 2 now contains $Z\id$. Apply another single-qubit
operation to bring this to $X\id$, and (as in the above cases)
perform a CNOT with column 1 as control and column 2 as target.
\end{itemize}
\hrulefill \caption{Required operations to eliminate all Pauli
operators from column 2 of the stabiliser array shown above, in
the various cases encountered.\label{tab:xzr}}
\end{table}
\begin{table}
\hrulefill \\
Initial stabiliser array:
$$
\left(
\begin{array}{c}
X \\
Z \\
.
\end{array}
\right)
$$
Depending on the content of row 3, do the following:
\begin{itemize}
\item[$X$:] Multiply row 1 with row 3.
\item[$Y$:] Multiply row 1 with row 3, and then row 2 with row 3.
\item[$Z$:] Multiply row 2 with row 3.
\end{itemize}
\hrulefill
\caption{Required operations to eliminate any Pauli
operator from row 3 of the stabiliser array shown
above.\label{tab:xzc}}
\end{table}

Algorithm CNFP brings only that part of the stabiliser array in
normal form that belongs to party A. Nevertheless, this is enough
to read off the number of EPR pairs in the full reduction. This
will be proven below. The number of EPR pairs $p$ is simply given
by the number of $XZ$ pairs in the normal form of party A.
\subsection{Proof of part i)}
By suitable modification of the Proof of algorithm CNF1, it can be
shown that algorithm CNFP brings that part of the stabiliser array
belonging to party A to the form as shown in (\ref{eq:cnf2}), the
columns left of the double vertical line.

We next show that by further applying suitable column operations
on the columns of party B, the complete normal form of
(\ref{eq:cnf2}) can be obtained.

Consider the first $XZ$ pair in party A. By commutativity of the
generators, there must at least be 1 anticommuting pair on the
same rows in party B. By a column permutation and a suitable
single-qubit rotation, this anticommuting pair can be moved to the
first column of party B and be brought in $XZ$ form. Using
suitable CNOTs (see Table \ref{tab:xzr}) the operators right of
the $XZ$ pair can all be brought to an $\id$ operator. Again by
commutativity, the operators below the $XZ$ pair must then
automatically be all $\id$ operators. Indeed, if a row (below the
second) contained a Pauli operator in the first column of party B,
it would not commute with either the first row, the second, or
both.

One can proceed in a similar fashion with the second of party A's
$XZ$ pairs and party B's second column and third and fourth row,
and so forth until all of A's $XZ$ pairs have been treated in this
way.

What remains then are the rows below the horizontal line in
(\ref{eq:cnf2}). To show that the lower right block of party $B$
in (\ref{eq:cnf2}) can be brought to the form as advertised (i.e.\
containing only $X$ and $\id$ operators, as denoted by the
asterisks), we note that party A contains no anticommuting pairs
in those rows. Hence, the subarray consisting of party B's lower
right block (restricted to that block's columns) consists of
mutually commuting generators. By applying algorithm CNF1 to that
subarray it can be brought in single-party normal form, consisting
of $X$ and $\id$ operators only. Evidently, the row operations
performed by the CNF1 algorithm (row permutation and
multiplication) will also affect the corresponding rows in party
A. However, as party A has only $X$ and $\id$ operators in those
rows, no $Y$ or $Z$ operators will be introduced, and the end
result will also contain only $X$ and $\id$ operators. \qed
\section{Conclusion}
The stabiliser formalism is a convenient tool for the study of
entanglement properties of large quantum many-body systems. While
the stabiliser formalism provides an efficient description of the
quantum state in terms of eigenvalue equations, it is not
immediately obvious how to obtain physical properties directly
from these eigenvalue equations without explicitly having to write
out the corresponding quantum state. In this paper we have
presented, employing elementary tools, a number of normal forms
for pure {\em and} mixed stabiliser states. We have furthermore
provided explicit, detailed descriptions of algorithms, whose
convergence we have proven, that allow the generation of these
normal forms. Using these normal forms, we can compute {\em any}
entanglement measure, overlaps between stabiliser states and
various other quantities. Detailed descriptions of the algorithms
are provided that should make it straightforward to implement them
in any programming language and we are able to provide MatLab
suite of programs on request.

These algorithms provide a firm basis for the exploration
entanglement properties of stabiliser states and suitable
generalisations in a great variety of contexts. For example, it is
readily seen that our approach is suitable for the efficient
simulation of systems where the initial state is a linear
combination of a polynomial number of stabiliser states. This and
other applications will be explored in forthcoming publications.

\begin{acknowledgments}
We would like to thank H.J. Briegel, O. Dahlsten and J. Oppenheim
for discussions. This work is part of the QIP-IRC (www.qipirc.org)
supported by EPSRC (GR/S82176/0) and is also supported by the EU
Thematic Network QUPRODIS (IST-2001-38877) and the Leverhulme
Trust grant F/07 058/U.
Finally, we thank the anonymous referees for comments that substantially
improved the presentation, and also for pointing out a serious omission in our
first treatment of overlap and Uhlmann fidelity.
\end{acknowledgments}

\end{document}